\documentclass[letterpaper]{emulateapj} 
\usepackage{apjfonts}                   

\usepackage{epsfig,graphicx,latexsym,amsmath,amssymb}
\usepackage{natbib}
\usepackage{hyperref}
\usepackage{mathrsfs}
\usepackage{color}
\usepackage{url}

\newcommand{\peri}{\hat{d}}
\newcommand{\Msun}{\ensuremath{\rm M_\odot}}
\newcommand{\Mmbh}{\ensuremath{M_{\rm MBH}}}
\newcommand{\tlife}{\ensuremath{t_{\rm life}}}
\newcommand{\trelax}{\ensuremath{t_{\rm rel}}}

\newcommand{\Out}[1]{}
\newcommand{\scale}[3]{{\left(\frac{#1}{#2}\right)^{#3}}}

\bibpunct[,]{(}{)}{;}{a}{}{,}

\begin{document}

\author{
Pau Amaro-Seoane\altaffilmark{1,2}\thanks{e-mail: Pau.Amaro-Seoane@aei.mpg.de} \&
Marc Dewi Freitag\altaffilmark{3,4}\thanks{e-mail: Marc.Freitag@gmail.com}
}

\altaffiltext{1}{Max Planck Institut f\"ur Gravitationsphysik
(Albert-Einstein-Institut), D-14476 Potsdam, Germany}
\altaffiltext{2}{Institut de Ci{\`e}ncies de l'Espai (CSIC-IEEC), Campus UAB,
Torre C-5, parells, $2^{\rm na}$ planta, ES-08193, Bellaterra,
Barcelona, Spain}
\altaffiltext{3}{Institute of Astronomy, Madingley Road, Cambridge, CB3 0HA, UK}
\altaffiltext{4}{Gymnase de Nyon, Route de Divonne 8, 1260 Nyon, Switzerland}

\date{\today}

\title{Relativistic encounters in dense stellar systems}

\begin{abstract}
Two coalescing black holes (BHs) represent a conspicuous source of
gravitational waves (GWs). The merger involves 17 parameters in the general
case of Kerr BHs, so that a successful identification and parameter extraction
of the information encoded in the waves will provide us with a detailed
description of the physics of BHs. A search based on matched-filtering for
characterization and parameter extraction requires the development of some
$10^{15}$ waveforms. If a third additional BH perturbed the system,
the waveforms would not be applicable, and we would need to increase the number 
of templates required for a valid detection.
In this letter, we calculate the probability that more than two 
BHs interact in the regime of strong relativity in a dense stellar 
cluster.
We determine the physical properties necessary in a stellar system for three black holes
to have a close encounter in this regime and also for an existing binary of two BHs 
to have a strong interaction with a third
hole.
In both cases the event rate is negligible.
While dense stellar systems such as galactic nuclei, globular clusters and
nuclear stellar clusters are the breeding grounds for the sources of
gravitational waves that ground-based detectors like Advanced
LIGO and Advanced VIRGO will be exploring, the analysis of the waveforms in full general
relativity needs only to evaluate the two-body problem. This 
reduces the number of templates of waveforms to create by orders of magnitude.
\end{abstract}

\maketitle

\section{Introduction}
\label{sec.introduction}

The detection of merging black holes (BHs) is the holy grail of ground-based 
detectors of gravitational waves (GWs) such
as LIGO and VIRGO. For the search for GWs of compact binaries, the availability of
accurate waveform models for the full merger is crucial. Thanks to the success
of numerical relativity in simulating the late inspiral, merger and ringdown of
a binary of two BHs \citep{Pretorius:2005gq,Campanelli:2005dd,Baker05a}, we are
now able to perform a search with realistic templates. The conjunction of
post-Newtonian modelling of the inspiral phase and full numerical relativistic
simulations of the merger and ringdown is now a reality for comparable-mass
scenarios of mass ratios up to about 10 \citep[see
e.g.][]{Buonanno:1998gg,Buonanno:2007pf,Ajith:2009bn,SantamariaEtAl10}.
Nevertheless, the high cost of full numerical relativity simulations constitutes
a serious limitation to the development of waveforms.

It has been estimated that we need a bank of $10^{13}$ waveforms for the
identification of a binary of two BHs with an F-statistic based grid
search at a {\em minimal match} \citep{CornishPorter07}.  On the other hand,
since the BHs are most likely Kerr, the number goes up to $\sim 10^{15}$ to
cover the spin parameter. In the case of stellar-mass BHs, since we cannot
detect the sky-location of the source, we are limited to a two-dimensional
space, the masses of the BHs, so that the number is considerably reduced; we
need to create about $\sim 10^5 - 10^6$ templates.  This is the reason why in
the last years there has been a significant effort in developing alternative
approaches, such as Monte-Carlo schemes, genetic algorithms,
Metropolis-Hastings methods and Nested Sampling techniques
\citep{CornishPorter06,LangHughes06,GairPorter09,PetiteauEtAl09}.

\cite{CampanelliEtAl08} addressed for the fist time fully relativistic
long-term numerical evolutions of three equal-mass BHs and found that the
merger dynamics is very distinct from binaries. In particular, the trajectories
were intricate and led to singular waveforms, as e.g. their figure 4 shows, in
which we can see two mergers.  Recently there has been an effort in calculating
in detail the waveforms of systems of three and four BHs interacting in full
GR. \cite{GalavizEtAl10} have developed a knowledgeable scheme to study the
waveforms of such configurations and find intricate templates for the waves.
Also, \cite{JaramilloLousto10} have addressed the problem of critical BH
separations for the formation of a common apparent horizon. The authors study
in detail the aligned equal mass cases for up to 5 BHs.

If we increase the number of BHs involved in the GW, the number of templates to
develop increases enormously. 
Putting it in Neil Cornish' words, ``The sensitive dependence on
initial conditions will send the template count through the roof''.
It is consequently important
to understand the limits imposed by the physical systems which harbour these
sources of GWs. Therefore we address the question of the existence of a system
with more than two BHs in a relativistic regime.  In section \ref{sec.two_bhs}
we calculate the probability of having a relativistic three-body encounter in a
dense stellar cluster with three BHs initially unbound.  In section
\ref{sec.bin_bh} we estimate the possibility that an already formed binary of
two BHs interacts relativistically with a third BH. In section
\ref{sec.discussion} we summarise our results and give the conclusions.

\section{Three-body relativistic encounters of unbound BHs}
\label{sec.two_bhs}

Rough estimtes will be sufficient to show how unlikely triple relativistic
encounters are.  Therefore, for simplicity, we assume that all BHs in a given
stellar system have the same mass, $m$.  Let us assume that we have two of them
fly by with a periapsis distance of a few Schwarzschild radii, $R_{\rm peri} =
\peri\, Gm/c^2$. Therefore, $\peri$ is the periapsis distance in units of
$Gm/c^2$, with $G$ the gravitational constant and $c$ the speed of light.  If
the relative velocity between the two objects at large separation is $V_{\rm
rel}$, the cross-section for an encounter with periapsis distance within
$\peri$ can be estimated as

\begin{equation}
S  = \pi \peri^2 \frac{G^2m^2}{c^4}
     \left[1+ \frac{2}{\peri}\left(\frac{c}{V_{\rm rel}}\right)^2 \right] \approx 2\pi\peri \frac{G^2m^2}{c^2\sigma^2}.
\label{eq.cross_sec_2BHs}
\end{equation}

\noindent
We have assumed that the BHs are in a non-relativistic stellar cluster,
as observed, $V_{\rm rel}\approx \sigma \ll c$. Here, $\sigma$ is the
3D velocity dispersion in the cluster, defined such that the total kinectic
energy in the cluster is $\frac{1}{2}M\sigma^2$, where $M$ is the total
mass of the cluster.

Hence, the encounter rate for one BH is 
$1/t_{\rm enc}=n\,S\,V_{\rm rel} \cong 2\pi\peri G^2m^2n/(c^2\sigma)$, where $n$ is the number density of BHs.
In order to obtain the rate of relativistic three-body encounters, one multiplies
by the probability of having a third object within a volume $\sim R^3_{\rm peri}$,
$P_{3}=n\,R^3_{\rm peri}$ \citep[see p. 201 of][]{HeggieHut03}

\begin{equation}
\frac{1}{t_{3}} \cong P_{3}\,\frac{1}{t_{\rm enc}}\cong n \peri^3\frac{G^3m^3}{c^6}2\pi\peri
                      \frac{G^2m^2n}{c^2\sigma}
                \cong 2\pi\peri^4\frac{G^5m^5n^2}{c^8\sigma}.
\end{equation}

\noindent
This is the rate per object. We now calculate the rate for a whole cluster of $N$ BHs,

\begin{equation}
\frac{1}{t_{3,\,{\rm tot}}} = \frac{N}{t_{3}} \cong
                           2\pi\peri^4 \frac{G^5m^5N^3}{c^8\sigma R^6}
                           =2\pi\peri^4 \frac{G^5M^5}{c^8\sigma R^6N^2},
\end{equation}
with $M=Nm$ and $n\approx N/R^3$. We ascertain now that the cluster is
self-gravitating, so that $GM/R \cong \sigma^2$, where $R$ is a typical
length-scale of the cluster, for instance its half-mass radius, i.e.\ the
radius of sphere containg one half of the BHs \citep[see for instance][]{BinneyTremaine08}

\Out{
\begin{equation}
\frac{1}{t_{3,\,{\rm tot}}} \cong 2\pi\peri^4 \left(\frac{\sigma}{c}\right)^9 \frac{c}{RN^2}
                           = 2\pi\peri^4 \left(\frac{\sigma}{c}\right)^8 \frac{1}{N^2}\frac{1}{t_{\rm cross}}.
\end{equation}

\noindent
In the last equation we have employed the inverse of the crossing time,
$t^{-1}_{\rm cross}=\sigma/R\cong\sqrt{G\rho} \cong \sigma^3/(GM)$. Therefore,

\begin{equation}
\frac{t_{\rm cross}}{t_{3,\,{\rm tot}}} \cong 9.63 \cdot 10^{-36}  
                                                   \left(\frac{\peri}{10}\right)^4
                                                   \left(\frac{\sigma}{100\,{\rm km\,s}^{-1}}\right)^8
                                                   \left(\frac{N}{10^6}\right)^{-2}.
\label{eq.T3_Tcross}
\end{equation}
}

\begin{equation}
 \begin{split}
 \frac{1}{t_{3,\,{\rm tot}}} & \cong 2\pi\peri^4\frac{\left(Gm\right)^{9/2}N^{5/2}}{c^8R^{11/2}} \\
 & \cong 2\cdot10^{-18}\,{\rm Gyr}^{-1}\times \\
 & \quad \scale{\peri}{100}{4}\scale{m}{10\,\Msun}{\frac{9}{2}}\scale{N}{10^6}{\frac{5}{2}}\scale{R}{0.1\,{\rm pc}}{-\frac{11}{2}}.
 \end{split}
 \label{eq.T3_num}
\end{equation}

\noindent

This equation makes it clear that the probability of even just one triple
encounter in a relativistic regime (small value of $\peri$) is extremely low,
even if one manages to pack a million stellar BHs within a sphere with a radius
of $0.1\,\rm pc$.

So far, we have considered only a self-gravitating cluster of compact objects
(COs), such as the stellar BHs, that have accumulated at the centre of a
globular cluster as a result of mass segregation
\citep{BHMMcMPZ03,BHMMcMPZ03b,BaumgardtEtAl04b}. However, a more promising
environment for the accumulation of a large number of COs in a small volume is
the centre of a galactic nucleus. There the gravitational force is likely
dominated by a massive black hole with a mass $\Mmbh\gtrsim 10^5\,\Msun$
\citep[][]{KR95,Magorrian98,GuelketinEtAl09}. In that
case, $\sigma \cong \sqrt{G\Mmbh/R}$, and equation~\ref{eq.T3_num} must be
modified with a factor $\sqrt{M/\Mmbh}$. This shows that relativistic triple
interactions are even less probable within COs orbiting MBHs than in a
self-gravitating cluster.

The lifetime of an isolated self-gravitating cluster is limited by 2-body
relaxation, i.e.\ the exchange of energy between stars during hyperbolic 2-body
encounters. Relaxation drives the overall expansion and evaporation of the
cluster \citep[see for instance][]{HeggieHut03}\footnote{Energetically, the
overall cluster expansion can occur only because a very small number of stars,
at the centre of the cluster, are getting more and more bound together. A
single tight binary can drive this expansion by releasing energy to passing-by
objects until it is ejected from the cluster or merges. In the next section, we
consider whether relativistic encounters can occur during interaction between a
binary and a single object.}.  We set an optimistic upper bound for the
lifetime of the cluster, $\tlife\lesssim 100\,\trelax$. The relaxation time is
approximately \citep{BinneyTremaine08}

\begin{equation}
\trelax\cong \frac{N}{\ln(0.1\,N)}\frac{R}{\sigma}.
\end{equation}

\noindent
Hence, we can estimate the total number of relativistic 3-body encounters over the lifetime of a self-gravitating cluster,

\begin{equation}
\begin{split}
N_{3,\,\rm singles} & \cong \frac{\tlife}{t_{3,\,{\rm tot}}}\lesssim 2\pi\frac{100}{\ln(0.1\,N)}\peri^4\frac{G^4m^4N^3}{c^8R^4}\\
& \cong 3\cdot10^{-18}\,\times \\
 & \quad \scale{\peri}{100}{4}\scale{m}{10\,\Msun}{4}\scale{N}{10^6}{3}\scale{R}{0.1\,{\rm pc}}{-4}.
\label{eq.N3sgl}
\end{split}
\end{equation}

\section{Three-body interactions between a binary and a single BH}
\label{sec.bin_bh}

The possibility of a relativistic encounter between three single objects being
ruled out, we turn to the possibility of achieving a 3-body relativistic
interaction through the encounter between a binary and a single object. Here
the difficulty is that the binary itself needs to be already in a relativistic
regime. This implies that it is emitting gravitational radiation at a high
rate, hence his lifetime is necessarily very limited.  Consider a binary of two
BHs of semi-major axis $a$ and masses $m_1$ and $m_2$ and a single BH of mass $m_3$ passing at a distance of $d$ from the center of mass of the binary. \Out{The orbital period is $t_{\rm orb}
\cong 1/(\sqrt{{G\,(m_1+m_2)}/{a^3}})$}

Approximately, the binary will need a time $t_{\rm GW}$ to shrink its orbit due
to the emission of gravitational waves as estimated by \cite{Peters64},

\begin{equation}
 t_{\rm GW} = \frac{5}{256}\frac{a^4c^5}{G^3\,m_1m_2\,(m_1+m_2)} 
\label{eq.tgw}
\end{equation}

Let us now estimate the timescale $t_3$ for the system to interact with a third
black hole that flies by at a distance of $\leq d$ of the centre-of-mass of the
binary. If the mass of the third BH is $m_3$ and the relative velocity to the
binary $V_{\rm rel}$, the cross section taking gravitational focusing into
account is

\begin{equation}
S = \pi d^2 \left[1+\frac{G\,(m_1+m_2+m_3)}{V_{\rm rel}^2\,d} \right] 
\cong \pi d \frac{G\,(m_1+m_2+m_3)}{V_{\rm rel}^2}
\label{eq.crosssec}
\end{equation}

Therefore, the timescale is

\begin{equation}
t_3  =     \frac{1}{n V_{\rm rel} S} \\
    \cong \frac{V_{\rm rel}}{n \pi d G (m_1+m_2+m_3)}.
\label{eq.t3}
\end{equation}

\noindent
\Out{In the last equation $n$ is the numerical density. The interaction duration is of
order $t_{\rm inter} \approx d/V_{\rm rel}$. }We assume that the three objects have a 
similar mass $m$; i.e. $m_1 \approx m_2 \approx m_3 := m$. Hence,

\begin{equation}
t_{\rm GW}  \cong \frac{5}{512}\frac{c^5\,a^4}{G^3m^3},\mbox{\ \ }
t_3         \cong \frac{V_{\rm rel}}{3 \pi n d G m}.
\end{equation}

We again measure distances in units of $GM/c^2$, with $\hat{a} := a/(G m/c^2)$
and $\peri := d/(G m/c^2)$. Typical interesting values are
$\hat{a}\approx{\rm a~few}$ and $\peri\approx{\rm few~tens}$. Therefore,

\begin{equation}
t_{\rm GW} \cong
\frac{1}{100}\frac{Gm}{c^3}\hat{a}^4
\end{equation}

\noindent
and

\begin{equation}
t_3 \cong \frac{1}{10}\frac{V_{\rm rel}c^2}{n\peri(Gm)^2}.          
\end{equation}

\noindent
\Out{In the last equation we have defined $\hat{n}:=n(Gm/c^2)^3$ for convenience.}
We then obtain the ratio of the two timescales,

\begin{equation}
\begin{split}
\frac{t_{\rm GW}}{t_3} & \cong \frac{1}{10}\frac{G^3m^3n\peri\hat{a}^4}{c^5 V_{\rm rel}}\\
& \cong 3\cdot 10^{-14} \times\\
&\quad \scale{V_{\rm rel}}{10\,{\rm km}\,{\rm s}^{-1}}{-1}\scale{\hat{a}^4\peri}{(100)^5}{}\scale{m}{10\,\Msun}{3}\scale{n}{10^{10}\,{\rm pc}^{-3}}{}.   
\end{split}
\label{eq.ratio}
\end{equation}

For a self-gravitating cluster, we can rewrite this in terms of $N$, and $R$, using $V_{\rm rel}\approx \sigma$,

\begin{equation}
\begin{split}
\frac{t_{\rm GW}}{t_3} & \cong \frac{1}{10}\frac{(Gm)^{5/2}\peri\hat{a}^4 N^{1/2}}{c^5 R^{5/2}}\\
& \cong 5\cdot 10^{-17} \times\\
&\quad \scale{\hat{a}^4\peri}{(100)^5}{}\scale{m}{10\,\Msun}{\frac{5}{2}}\scale{N}{10^6}{\frac{1}{2}}\scale{R}{0.1\,{\rm pc}}{-\frac{5}{2}}.   
\end{split}
\label{eq.ratio2}
\end{equation}

\noindent
This quantity can be interpreted as the probability that a relativistic binary
has a close encounter with a single object before it merges. The total number
of mergers that can occur in the evolution of a cluster of $N$ objects is
$N_{\rm merg}\le N$,  with $N_{\rm merg}= N$ only possible if all objects merge
together \citep[a scenario to form a MBH from a cluster of stellar BHs, see][]{Lee93,KupiEtAl06}.
Hence, the total number of relativistic single-binary interactions over the
lifetime of a cluster (evolving through a succession of binary mergers) is

\begin{equation}
\begin{split}
N_{3,\,\rm bin} & \lesssim N\frac{t_{\rm GW}}{t_3}
\cong 5\cdot 10^{-11} \times\\
&\quad \scale{\hat{a}^4\peri}{(100)^5}{}\scale{m}{10\,\Msun}{\frac{5}{2}}\scale{N}{10^6}{\frac{3}{2}}\scale{R}{0.1\,{\rm pc}}{-\frac{5}{2}}.   
\label{eq.N3bin}
\end{split}
\end{equation}

\Out{
If we require that $t_{\rm GW} > t_3$, we then need a normalized numerical density of

\begin{equation}
\hat{n} > 10 \left(\frac{V_{\rm rel}}{c}\right)\frac{1}{\hat{a}^4\peri}  
\end{equation}

Say that e.g. $\hat{a}=5$, $\peri=30$ and $V_{\rm rel} = 30\,{\rm
km\,s}^{-1}$.  We have that $V_{\rm rel}/c = 10^{-4}$ and $n \gtrsim
10^{-7}\left[{c^2}/({Gm})\right]^3$. For $m=10\Msun$, $Gm/c^2 \approx
30\,{\rm km}=10^{-12}\,{\rm pc}$. Hence we see that we need a stellar numerical
density of $n \gtrsim 10^{29}\,{\rm pc}^{-3}$; resultantly $\rho \gtrsim
10^{30}\Msun/{\rm pc}^3$. This value is obviously too large to exist in
nature, even for galactic nuclei.  The typical densities observed may exceed
the core density of globular clusters by orders of magnitude. Our own Galactic
Center has a density of $\sim 10^7-10^8 \Msun~{\rm pc}^{-3}$.

If we had such high densities, we can estimate what would be the mass of the
cluster in BHs,

\begin{equation}
\rho_{\bullet} = \frac{M_{\bullet}}{\left(GM_{\bullet}/c^2\right)^3}=
                 \frac{c^6}{G^3M_{\bullet}^2}
\end{equation}

The value of $M_{\bullet}$ for which $\rho_{\bullet} = \rho$ is

\begin{equation}
\rho_{\bullet} = \left(\frac{M_{\bullet}}{\Msun}\right)^{-2} 
                 \times 10^{36} \Msun\,{\rm pc}^{-3}
\end{equation}

Thus, we can see that $\left(M_{\bullet}/\Msun\right)^2=10^6$, which 
implies that $M_{\bullet}=10^3 \Msun$
}

\section{Discussion}
\label{sec.discussion}

In this work we have addressed the formation of systems of three BHs in the
strong gravity regime. For that we have first studied the probability that
three BHs interact in a dense stellar cluster and we conclude that it is
totally negligible.  We have then addressed the possibility that a binary of
BHs which has previously formed in the cluster interacts relativistically with
a third BH. We judge that the stellar system harbouring the BHs needs to have
unachievable densities.

\begin{figure*}
  \begin{center}
    \begin{minipage}[t]{0.58\linewidth}
    \includegraphics[scale=0.42,angle=270,clip]{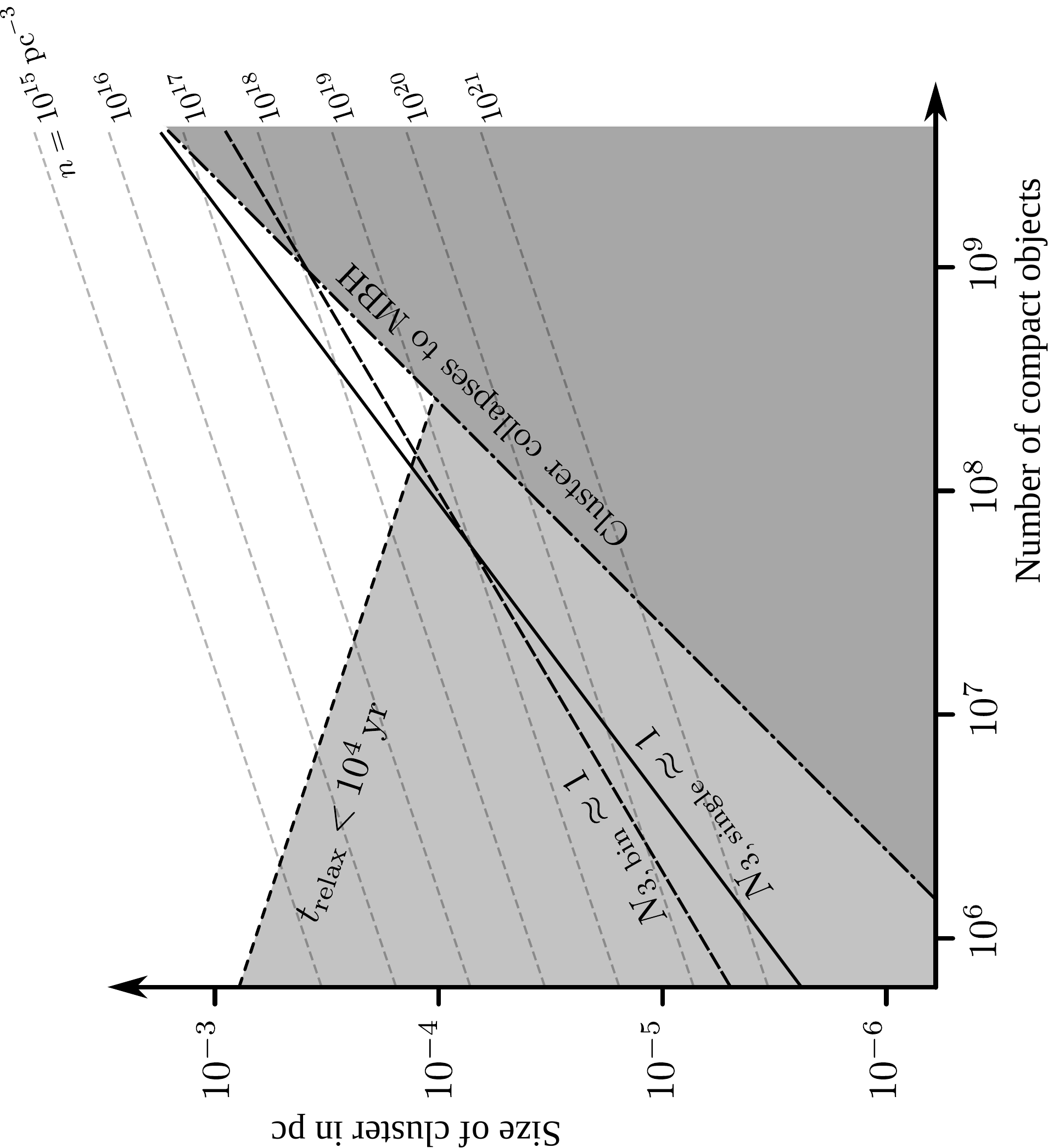}
    \end{minipage}\hfill
    \begin{minipage}[t]{0.38\linewidth}
      \caption{
Parameter space for the occurrence of triple relativistic encounters. For this
figure, we consider a self-gravitating cluster made of a given number of
identical compact object with an individual mass of $10\,\Msun$. The line
labeled $N_{3,\,\rm bin}\approx 1$ is based on equation~\ref{eq.N3bin} with
$\peri=\hat{a}=10^4$. Below this line, in our simple estimate, at least one
binary--single encounter in which the three objects are relativistic should
occur during the life-time of the cluster (which is limited by successive
mergers, see text). The line labeled $N_{3,\,\rm single}\approx 1$ is based on
equation~\ref{eq.N3sgl} with $\peri=10^4$. Below this line, one can expect at
least one triple relativistic encounter between single objects to occur during
the life-time of the cluster (limited by 2-body relaxational processes). The
large gray area in the bottom-right section is excluded from the parameter
space because the whole cluster would be smaller than its Schwarzschild radius,
i.e.\ it would immediately collapse into a single massive black hole. The other
gray region correspond to clusters in which the 2-body relaxation time would be
shorter than $10^4$ years. Such a cluster would probably not last long enough
for stars to form and turn into compact objects. The dashed diagonal lines in
gray indicate an estimate of the number density in the cluster, from the
relation $n\approx N/R^3$
               \label{fig.NR}}
    \end{minipage}
  \end{center}
\end{figure*}

Based on simple physical arguments, we have established that the time scales
for a triple relativistic encounter to occur in a cluster of stellar-mass
compact objects is extremely long.  We recall that we consider two extreme
cases. In the first situation, we neglect the existence (and formation) of
binaries. Hence, we consider that three single objects have to find themselves,
by chance, within a few (tens of) Schwarzschild radii of each other. In this
case, we have admitted that the cluster cannot survive (with a high stellar
density) for more than about 100 relaxation time. Indeed, in such a long time
scale, most of the cluster would expand to lower and lower densities and a very
significant fraction of the object would escape, with only a very small number
of objects getting closer and closer to provide energy for this evolution
\citep{HeggieHut03}. The second (much idealized) case is that of a cluster made mostly of
binaries. In that case, one can hope that it would suffice for an object to
interact closely with a binary but in order for the interaction to be
relativistic for the three objects, the binary must be so tight that its
lifetime is limited by emission of gravitational waves. Accordingly, we
consider that the lifetime of the whole cluster is limited by the successive
merger of binaries. Any real cluster would present a situation which is
somewhat in between these two extremes. In particular, the evolution of a cluster
made of single objects would naturally lead to the formation of binaries during
core collapse \citep[see e.g.][]{HeggieHut03}. We stress that we have made a large number of
simplifications in our estimates but always in such a way as to overestimate
the rate of triple relativistic events. For instance, we have assumed that all
the binaries in a cluster are relativistic. This limits their lifetime but
non-relativistic binaries are useless for our purpose.

With figure~\ref{fig.NR}, we can estimate what conditions are required for at
least one such encounter to take place during the lifetime of the cluster. For
this figure, we have assumed $m=10\,\Msun$ and $\peri=\hat{a}=10^4$. The latter
values correspond to encounters that are only weakly relativistic. Even with
such values, the figure shows that most of the parameter space for $N_{3,\,\rm
single}\gtrsim 1$ or $N_{3,\,\rm bin}\gtrsim 1$ is excluded, either because the
cluster, as a whole, would be a massive black hole (for large $N$ and small
$R$) or because the cluster would have such a small 2-body relaxation time that
it wouldn't have time to form (for smaller $N$ and small $R$). Indeed, the
evolution of massive stars into compact objects (neutron stars or black holes)
requires at least 3~million years 
. Because of these constraints, it
appears that clusters hosting triple relativistic encounters have to be made of
at least $10^8$ compact objects concentrated within a region smaller than
$10^{-3}\,{\rm pc}$. In fact, for $N\approx10^8$, the size has to be of order
$10^{-4}\,{\rm pc}$. The corresponding number density is comprised between $10^{17}$ and
$10^{19}\,{\rm pc}^{-3}$.

Such values are beyond observed ones by many orders of magnitude. For instance,
the stellar density in Galactic globular clusters is, at most, of the order of
$10^{5}\,{\rm pc}^{-3}$ \citep[see][and the 2003 on-line update
\url{http://www.physics.mcmaster.ca/~harris/mwgc.dat}]{Harris96}. Already the
necessary number of compact objects is much larger than what one can expect in
the kind of clusters that exist. A globular cluster might contain up to $10^7$
stars but only a very small fraction of them would become BHs. This number
fraction $f$ depends on the initial mass function (IMF) of high mass stars.
Some galactic nuclei seem to be top-heavy \citep{ManessEtAl07}, so that in
principle $f$ ranges between $10^{-3}$ and $10^{-2}$, yielding, at most,
$N=10^5$ BHs in very large globular cluster.

A galaxy contains a much larger number of compact objects but most of them
inhabit regions of very low density. BHs born in the central regions are likely
to accumulate at the centre, in the galactic nucleus, through the process of
mass segregation, which is basically an effect of dynamical friction. But, in
the case of our Galaxy, dynamical friction is only effective, over a Hubble
time, within a few parsecs of the centre. Hence, at most of order $10^4$ BHs
may have gathered within the innermost $0.3\,{\rm pc}$
\citep{FAK06a,Amaro-SeoaneFreitagPreto10}. A dark mass concentration weighing
$4\times 10^6\,\Msun$ has been detected at the Galactic centre through the
analysis of the orbits of bright IR stars, the so-called S- or SO-stars,
around the weak radio source Sgr\,A$^\ast$
\citep{EisenhauerEtAl05,GhezEtAl05,GhezEtAl08,GillessenEtAl09,GenzelEtAl10}. It
is generally assumed that it is a massive black hole although the only strict
constraint on its size is that it has to fit within the periapse of the IR
stars' orbits, the tightest of which is that of S-2 (SO-2), imposing $R\lesssim
5\times10^{-4}$\,pc. If we assume that this object is actually a cluster of
$10\,\Msun$ BHs, with $N\approx 4\times 10^5$, it would need to be so compact
in order to host triple relativistic encounters that its relaxation time would
be extremely short, making its existence at the present time an extraordinary
coincidence \citep[See also][]{Maoz98,Miller06}. Furthermore, no mechanism to
form such a dense cluster is known. 
 
While we have limited the analysis to a system of three BHs, it is nevertheless
obvious that for more BHs the event rates are much more unlikely. We therefore
conclude that the waveforms to develop need only to include two BHs for the
searches of GWs in the data streams.

\acknowledgments 

PAS work was partially supported by the DLR (Deutsches Zentrum f\"ur Luft- und
Raumfahrt). We are thankful to Neil Cornish, Ed Porter and Antoine Petiteau for
discussions and to Luciano Rezzolla for his encouragement to publish this small
note.


\begin{thebibliography}{37}
\expandafter\ifx\csname natexlab\endcsname\relax\def\natexlab#1{#1}\fi

\bibitem[{{Ajith} {et~al.}(2009){Ajith}, {Hannam}, {Husa}, {Chen}, {Bruegmann},
  {Dorband}, {Mueller}, {Ohme}, {Pollney}, {Reisswig}, {Santamar\'ia}, \&
  {Seiler}}]{Ajith:2009bn}
{Ajith} P., {Hannam} M., {Husa} S., {Chen} Y., {Bruegmann} B., {Dorband} N.,
  {Mueller} D., {Ohme} F., {Pollney} D., {Reisswig} C., {Santamar\'ia} L.,
  {Seiler} J., 2009

\bibitem[{{Amaro-Seoane} {et~al.}(2010){Amaro-Seoane}, {Freitag}, \&
  {Preto}}]{Amaro-SeoaneFreitagPreto10}
{Amaro-Seoane} P., {Freitag} M., {Preto} M., 2010, To be submitted

\bibitem[{Baker {et~al.}(2006)Baker, Centrella, Choi, Koppitz, \& van
  Meter}]{Baker05a}
Baker J.~G., Centrella J., Choi D.-I., Koppitz M., van Meter J., 2006, Phys.
  Rev. Lett., 96, 111102

\bibitem[{{Baumgardt} {et~al.}(2003{\natexlab{a}}){Baumgardt}, {Hut}, {Makino},
  {McMillan}, \& {Portegies Zwart}}]{BHMMcMPZ03}
{Baumgardt} H., {Hut} P., {Makino} J., {McMillan} S., {Portegies Zwart} S.,
  2003{\natexlab{a}}, ApJ Lett., 582, L21

\bibitem[{{Baumgardt} {et~al.}(2004){Baumgardt}, {Makino}, \&
  {Ebisuzaki}}]{BaumgardtEtAl04b}
{Baumgardt} H., {Makino} J., {Ebisuzaki} T., 2004, ApJ, 613, 1143

\bibitem[{{Baumgardt} {et~al.}(2003{\natexlab{b}}){Baumgardt}, {Makino}, {Hut},
  {McMillan}, \& {Portegies Zwart}}]{BHMMcMPZ03b}
{Baumgardt} H., {Makino} J., {Hut} P., {McMillan} S., {Portegies Zwart} S.,
  2003{\natexlab{b}}, ApJ Lett., 589, L25

\bibitem[{{Binney} \& {Tremaine}(2008)}]{BinneyTremaine08}
{Binney} J., {Tremaine} S., 2008, {Galactic Dynamics: Second Edition}, {Binney}
  J., {Tremaine} S., eds. Princeton University Press

\bibitem[{Buonanno \& Damour(1999)}]{Buonanno:1998gg}
Buonanno A., Damour T., 1999, Phys. Rev. D, 59, 084006

\bibitem[{Buonanno {et~al.}(2007)}]{Buonanno:2007pf}
Buonanno A., {et~al.}, 2007, Phys. Rev., D76, 104049

\bibitem[{Campanelli {et~al.}(2006)Campanelli, Lousto, Marronetti, \&
  Zlochower}]{Campanelli:2005dd}
Campanelli M., Lousto C.~O., Marronetti P., Zlochower Y., 2006, Phys. Rev.
  Lett., 96, 111101

\bibitem[{{Campanelli} {et~al.}(2008){Campanelli}, {Lousto}, \&
  {Zlochower}}]{CampanelliEtAl08}
{Campanelli} M., {Lousto} C.~O., {Zlochower} Y., 2008, Physical Review D, 77,
  101501

\bibitem[{{Cornish} \& {Porter}(2006)}]{CornishPorter06}
{Cornish} N.~J., {Porter} E.~K., 2006, Classical and Quantum Gravity, 23, 761

\bibitem[{{Cornish} \& {Porter}(2007)}]{CornishPorter07}
---, 2007, Physical Review D, 75, 021301

\bibitem[{{Eisenhauer} {et~al.}(2005){Eisenhauer}, {Genzel}, {Alexander},
  {Abuter}, {Paumard}, {Ott}, {Gilbert}, {Gillessen}, {Horrobin}, {Trippe},
  {Bonnet}, {Dumas}, {Hubin}, {Kaufer}, {Kissler-Patig}, {Monnet},
  {Str{\"o}bele}, {Szeifert}, {Eckart}, {Sch{\"o}del}, \&
  {Zucker}}]{EisenhauerEtAl05}
{Eisenhauer} F., {Genzel} R., {Alexander} T., {Abuter} R., {Paumard} T., {Ott}
  T., {Gilbert} A., {Gillessen} S., {Horrobin} M., {Trippe} S., {Bonnet} H.,
  {Dumas} C., {Hubin} N., {Kaufer} A., {Kissler-Patig} M., {Monnet} G.,
  {Str{\"o}bele} S., {Szeifert} T., {Eckart} A., {Sch{\"o}del} R., {Zucker} S.,
  2005, ApJ, 628, 246

\bibitem[{{Freitag} {et~al.}(2006){Freitag}, {Amaro-Seoane}, \&
  {Kalogera}}]{FAK06a}
{Freitag} M., {Amaro-Seoane} P., {Kalogera} V., 2006, ApJ, 649, 91

\bibitem[{{Gair} \& {Porter}(2009)}]{GairPorter09}
{Gair} J.~R., {Porter} E.~K., 2009, ArXiv e-prints

\bibitem[{{Galaviz} {et~al.}(2010){Galaviz}, {Bruegmann}, \&
  {Cao}}]{GalavizEtAl10}
{Galaviz} P., {Bruegmann} B., {Cao} Z., 2010, ArXiv e-prints

\bibitem[{{Genzel} {et~al.}(2010){Genzel}, {Eisenhauer}, \&
  {Gillessen}}]{GenzelEtAl10}
{Genzel} R., {Eisenhauer} F., {Gillessen} S., 2010, ArXiv e-prints

\bibitem[{{Ghez} {et~al.}(2005){Ghez}, {Salim}, {Hornstein}, {Tanner}, {Lu},
  {Morris}, {Becklin}, \& {Duch{\^e}ne}}]{GhezEtAl05}
{Ghez} A.~M., {Salim} S., {Hornstein} S.~D., {Tanner} A., {Lu} J.~R., {Morris}
  M., {Becklin} E.~E., {Duch{\^e}ne} G., 2005, ApJ, 620, 744

\bibitem[{{Ghez} {et~al.}(2008){Ghez}, {Salim}, {Weinberg}, {Lu}, {Do}, {Dunn},
  {Matthews}, {Morris}, {Yelda}, {Becklin}, {Kremenek}, {Milosavljevic}, \&
  {Naiman}}]{GhezEtAl08}
{Ghez} A.~M., {Salim} S., {Weinberg} N.~N., {Lu} J.~R., {Do} T., {Dunn} J.~K.,
  {Matthews} K., {Morris} M.~R., {Yelda} S., {Becklin} E.~E., {Kremenek} T.,
  {Milosavljevic} M., {Naiman} J., 2008, ApJ, 689, 1044

\bibitem[{{Gillessen} {et~al.}(2009){Gillessen}, {Eisenhauer}, {Trippe},
  {Alexander}, {Genzel}, {Martins}, \& {Ott}}]{GillessenEtAl09}
{Gillessen} S., {Eisenhauer} F., {Trippe} S., {Alexander} T., {Genzel} R.,
  {Martins} F., {Ott} T., 2009, ApJ, 692, 1075

\bibitem[{{G{\"u}ltekin} {et~al.}(2009){G{\"u}ltekin}, {Richstone}, {Gebhardt},
  {Lauer}, {Tremaine}, {Aller}, {Bender}, {Dressler}, {Faber}, {Filippenko},
  {Green}, {Ho}, {Kormendy}, {Magorrian}, {Pinkney}, \&
  {Siopis}}]{GuelketinEtAl09}
{G{\"u}ltekin} K., {Richstone} D.~O., {Gebhardt} K., {Lauer} T.~R., {Tremaine}
  S., {Aller} M.~C., {Bender} R., {Dressler} A., {Faber} S.~M., {Filippenko}
  A.~V., {Green} R., {Ho} L.~C., {Kormendy} J., {Magorrian} J., {Pinkney} J.,
  {Siopis} C., 2009, ApJ, 698, 198

\bibitem[{{Harris}(1996)}]{Harris96}
{Harris} W.~E., 1996, AJ, 112, 1487

\bibitem[{{Heggie} \& {Hut}(2003)}]{HeggieHut03}
{Heggie} D., {Hut} P., 2003, {The Gravitational Million-Body Problem: A
  Multidisciplinary Approach to Star Cluster Dynamics}, {Heggie, D.~\& Hut,
  P.}, ed.

\bibitem[{{Jaramillo} \& {Lousto}(2010)}]{JaramilloLousto10}
{Jaramillo} G., {Lousto} C.~O., 2010, ArXiv e-prints

\bibitem[{{Kormendy} \& {Richstone}(1995)}]{KR95}
{Kormendy} J., {Richstone} D., 1995, ARA\&A, 33, 581

\bibitem[{{Kupi} {et~al.}(2006){Kupi}, {Amaro-Seoane}, \&
  {Spurzem}}]{KupiEtAl06}
{Kupi} G., {Amaro-Seoane} P., {Spurzem} R., 2006, MNRAS, L77+

\bibitem[{{Lang} \& {Hughes}(2006)}]{LangHughes06}
{Lang} R.~N., {Hughes} S.~A., 2006, Phys. Rev. D, 74, 122001

\bibitem[{{Lee}(1993)}]{Lee93}
{Lee} M.~H., 1993, ApJ, 418, 147

\bibitem[{{Magorrian} {et~al.}(1998){Magorrian}, {Tremaine}, {Richstone},
  {Bender}, {Bower}, {Dressler}, {Faber}, {Gebhardt}, {Green}, {Grillmair},
  {Kormendy}, \& {Lauer}}]{Magorrian98}
{Magorrian} J., {Tremaine} S., {Richstone} D., {Bender} R., {Bower} G.,
  {Dressler} A., {Faber} S.~M., {Gebhardt} K., {Green} R., {Grillmair} C.,
  {Kormendy} J., {Lauer} T., 1998, AJ, 115, 2285

\bibitem[{{Maness} {et~al.}(2007){Maness}, {Martins}, {Trippe}, {Genzel},
  {Graham}, {Sheehy}, {Salaris}, {Gillessen}, {Alexander}, {Paumard}, {Ott},
  {Abuter}, \& {Eisenhauer}}]{ManessEtAl07}
{Maness} H., {Martins} F., {Trippe} S., {Genzel} R., {Graham} J.~R., {Sheehy}
  C., {Salaris} M., {Gillessen} S., {Alexander} T., {Paumard} T., {Ott} T.,
  {Abuter} R., {Eisenhauer} F., 2007, ApJ, 669, 1024

\bibitem[{{Maoz}(1998)}]{Maoz98}
{Maoz} E., 1998, ApJ Lett., 494, 181

\bibitem[{{Miller}(2006)}]{Miller06}
{Miller} M.~C., 2006, MNRAS, 367, L32

\bibitem[{{Peters}(1964)}]{Peters64}
{Peters} P.~C., 1964, Physical Review, 136, 1224

\bibitem[{{Petiteau} {et~al.}(2009){Petiteau}, {Yu}, \&
  {Babak}}]{PetiteauEtAl09}
{Petiteau} A., {Yu} S., {Babak} S., 2009, ArXiv e-prints

\bibitem[{Pretorius(2005)}]{Pretorius:2005gq}
Pretorius F., 2005, Phys. Rev. Lett., 95, 121101

\bibitem[{{Santamar{\'{\i}}a} {et~al.}(2010){Santamar{\'{\i}}a}, {Ohme},
  {Ajith}, {Br{\"u}gmann}, {Dorband}, {Hannam}, {Husa}, {M{\"o}sta}, {Pollney},
  {Reisswig}, {Robinson}, {Seiler}, \& {Krishnan}}]{SantamariaEtAl10}
{Santamar{\'{\i}}a} L., {Ohme} F., {Ajith} P., {Br{\"u}gmann} B., {Dorband} N.,
  {Hannam} M., {Husa} S., {M{\"o}sta} P., {Pollney} D., {Reisswig} C.,
  {Robinson} E.~L., {Seiler} J., {Krishnan} B., 2010, Phys Rev D, 82, 064016

\end{thebibliography}
\end{document}